# Residual-CycleGAN based Camera Adaptation for Robust Diabetic Retinopathy Screening

Dalu Yang[1], Yehui Yang[1], Tiantian Huang, Binghong Wu, Lei Wang, Yanwu Xu[2]

Intelligent Health Unit, Baidu Inc, Beijing, China
`{yangdalu, yangyehui01, huangtiantian01, wubinghong, wanglei15, xuyanwu}@baidu.com`

**Abstract.** There are extensive researches focusing on automated diabetic retinopathy (DR) detection from fundus images. However, the accuracy drop is observed when applying these models in real-world DR screening, where the fundus camera brands are different from the ones used to capture the training images. How can we train a classification model on labeled fundus images acquired from only one camera brand, yet still achieves good performance on images taken by other brands of cameras? In this paper, we quantitatively verify the impact of fundus camera brands related domain shift on the performance of DR classification models, from an experimental perspective. Further, we propose camera-oriented residual-CycleGAN to mitigate the camera brand difference by domain adaptation and achieve increased classification performance on target camera images. Extensive ablation experiments on both the EyePACS dataset and a private dataset show that the camera brand difference can significantly impact the classification performance and prove that our proposed method can effectively improve the model performance on the target domain. We have inferred and labeled the camera brand for each image in the EyePACS dataset and will publicize the camera brand labels for further research on domain adaptation.

## 1 Introduction

Diabetic Retinopathy (DR) is a leading cause of blindness worldwide that affects approximately 30% of diabetes mellitus patients [1, 2]. There are extensive researches focusing on the automatic early detection of DR via color fundus photography, and they achieve remarkable classification performance [3, 4, 5, 6].

Despite that most of these high-performance results are reproducible with open-sourced code and specific public datasets, the generalizability of these researches is usually poor on fundus images with completely different distributions. As a result, the application of automatic DR detection in real-world DR screening programs is still limited. One of the key differences between experimental and real-world settings is the use of different brands of fundus cameras. It is common that the training dataset

---

[1] Equal contribution
[2] Corresponding author

does not contain images captured by the same camera model used in the DR screening site.

By collecting extra data or applying extra limitations, the impact of domain shift in real-world settings can be avoided. Gulshan et. al. collected fundus images from a large number of hospitals across the US and India, covering most mainstream camera brands and produce a model with good generalizability [3], but the data collection process is extremely costly. Abràmoff et. al. took a different approach by restricting the usage of the software on only one single camera model (TRC-NW400, Topcon) [7]. The data distribution is effectively under control, but this approach limits the application of automated DR screening to places that can afford the specific camera model.

On the other hand, without extra data collection or limitations, domain adaptation is the technique that can counter the domain shift and increase model performance on different data distributions. In recent years, deep learning based domain adaptation techniques have been well developed [8, 9], and applied widely on different medical imaging tasks [10, 11, 12]. However, domain adaptation is rarely addressed in fundus photography images. Currently [13] is the only work on fundus image domain adaptation, where the authors adapt the optic disc and optic cup segmentation model trained on images from Zeiss Visucam 500 to images from Canon CR-2. However, their experiments are performed on only one source domain and one target domain, thus fail to address the domain shift problem of different camera brands in general.

In this paper, we propose a camera adaptation method to address a general camera adaptation problem: the performance of DR screening models fluctuates across images from various fundus camera brands. The key contributions of this paper are: 1) To our knowledge, we are the first to quantitatively verify the impact of fundus camera brands related domain shift on the performance of deep learning DR classification models, from an experimental perspective. 2) We propose a novel camera-oriented residual-CycleGAN which can mitigate the domain shift and achieve increased classification performance on target camera images, without retraining the base model. We perform extensive ablation experiments on both public datasets and private datasets for comparison of the results. 3) We will publicize our camera brand grouping label of the whole EyePACS dataset to enable further adaptation researches. EyePACS is so far the largest fundus dataset of multiple camera brands. Our camera brand domain label on EyePACS will both help the development of real-world DR screening systems and benefit the researchers in the field of domain adaptation in general.

## 2    Methods

### 2.1    Overview

In this paper, we characterize the camera adaptation as an unsupervised domain adaptation problem, under the setting with only one labeled source domain (camera brand) and multiple unlabeled target domains. Formally, we have source domain $A$ with image set $X_A$ and label set $Y_A$, and a set of target domain image set $X_B$, $X_C$, $X_D$, etc. Our domain adaptation process consists of three steps, as shown in **Fig. 1**: 1) We first train a classifier *M* with $X_A$ and $Y_A$ (**Fig. 1**a); 2) We use $X_A$ and unlabeled $X_B$, $X_C$, $X_D$, etc. to obtain domain transformations $G_B$, $G_C$, $G_D$, etc., respectively (**Fig. 1**b); 3) We apply the transformations $G_B$, $G_C$, $G_D$ etc. on $x_B \in X_B$, $x_C \in X_C$, $x_D \in X_D$, etc. at test time, and use *M* to get the final classification results (**Fig. 1**c).

In general, *M* has a poor performance when applied directly on the target domain images $X_B$, $X_C$, $X_D$. With the domain transformation, although the classifier *M* is never trained on the target domain images, it has a better classification performance.

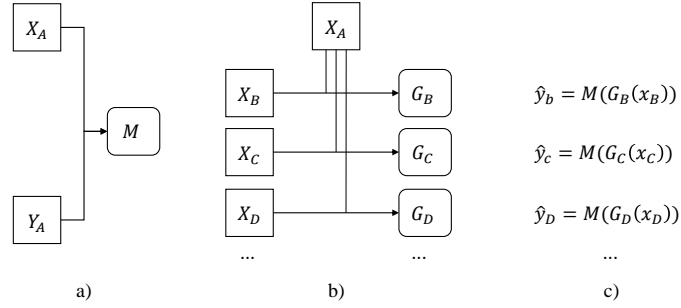

**Fig. 1.** Overview of the problem and domain adaptation process. $X_A$: source domain image set; $Y_A$: source domain label set; *M*: classifier trained with source domain images and labels; $X_B$, $X_C$, $X_D$: target domain image sets; $G_B$, $G_C$, $G_D$: domain transformation models; $\hat{y}_B$, $\hat{y}_C$, $\hat{y}_D$: predicted labels of target domain images $x_B$, $x_C$, $x_D$.

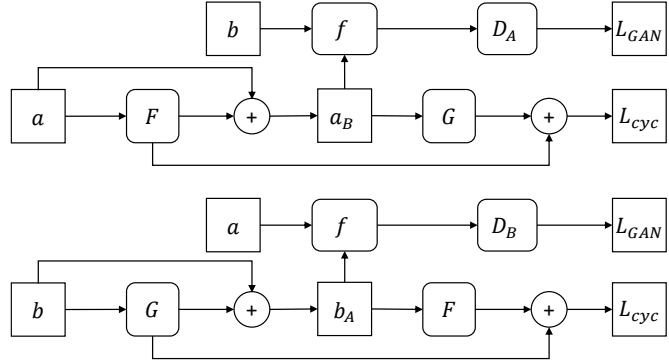

**Fig. 2.** The framework of residual-CycleGAN. *F* and *G*: residual generators applying on domain *A* and *B*; *a* and *b*: images from domain *A* and *B*; $a_B$ and $b_A$: transformed image *a* to domain *B* and image *b* to domain *A*; *f*: camera-oriented feature extractor; $D_B$ and $D_A$: discriminator for real domain *B* and *A*; $L_{cyc}$: cycle-consistency loss; $L_{GAN}$: generative-adversarial loss.

## 2.2 Camera oriented Residual-CycleGAN

The overall network resembles a CycleGAN-like structure, with residual generator and customized discriminator. **Fig. 2** illustrates our proposed domain transformation network. The major components of the network include: the residual generator *F* and *G*, the task specific feature extractor *f*, and the domain discriminator $D_A$ and $D_B$. For convenience, we denote the original images from domain A and B as *a* and *b*, and the transformed image as $a_B$ and $b_A$. The residual domain transformation is:

$$\begin{aligned} a_B &= a + F(a), \\ b_A &= b + G(b). \end{aligned} \quad (1)$$

Intuitively, the generator *F* takes image from domain A to a residual transformation, which serves as an additive filter that can be applied to *a* and result in a fake $a_B$ that has the distribution of domain B. Similarly, the generator *G* does the same thing for *b* to transform from domain B to domain A. In the remaining part of this section, we first formulate the domain transformation learning problem, then we present the details of each component in our network.

**Formulation of learning.** Following the vanilla CycleGAN formulation, our total objective function includes three components, i.e., the adversarial loss, the cycle-consistency loss, and the identity loss. The description of each loss is as follows:

1. *Adversarial Loss.* The adversarial loss is defined as:

$$\begin{aligned} \mathcal{L}_{GAN}(F, D_B) &= \mathbb{E}_{b \in B}[\log D_B(f(b))] + \mathbb{E}_{a \in A}\left[\log\left(1 - D_B(f(a_B))\right)\right], \\ \mathcal{L}_{GAN}(G, D_A) &= \mathbb{E}_{a \in A}[\log D_A(f(a))] + \mathbb{E}_{b \in B}\left[\log\left(1 - D_A(f(b_A))\right)\right]. \end{aligned} \quad (2)$$

   Notice $a_B$ and $b_A$ are the transformed images defined in (1). The adversarial loss regularizes the training through the minimax optimization, i.e., $\min_F \max_{D_B} \mathcal{L}_{GAN}(F, D_B)$ and $\min_G \max_{D_A} \mathcal{L}_{GAN}(G, D_A)$, to ensure that the transformed images are indistinguishable from the other domain.

2. *Cycle-consistency Loss.* Due to the residual nature of our network, the cycle-consistency loss can be expressed as:

$$\mathcal{L}_{cyc}(F, G) = \mathbb{E}[\|F(a) + G(a_B)\|_2] + \mathbb{E}[\|G(b) + F(b_A)\|_2]. \quad (3)$$

   Intuitively, minimizing this cycle-consistency loss encourages the sum of residues added to the original image to be zero, when the original image is transformed from domain *A* to *B* and back to *A*.

3. *Identity Loss.* The identity loss in our network takes the form of:

$$\mathcal{L}_{idt}(F, G) = \mathbb{E}[\|F(b)\|_2] + \mathbb{E}[\|G(a)\|_2]. \quad (4)$$

   The identity loss further encourages an identity mapping on real target domain image. [14] claims the application of identity loss prevents dramatic change of image hue, which is crucial when dealing with fundus images. We notice that in our experiment, using L2-norm in $\mathcal{L}_{cyc}$ and $\mathcal{L}_{idt}$ yields a faster convergence than using the L1-norm as stated in the original cycleGAN model.

Combining equation (2, 3, 4), the overall loss function is:

$$\begin{aligned} \mathcal{L}(F, G, D_A, D_B) &= \mathcal{L}_{GAN}(F, D_B) + \mathcal{L}_{GAN}(G, D_A) \\ &\quad + \lambda_1 \mathcal{L}_{idt}(F, G) + \lambda_2 \mathcal{L}_{cyc}(F, G). \end{aligned} \quad (5)$$

And the objective is to solve for *F* and *G*:

$$F^*, G^* = \arg\min_{F,G} \max_{D_A, D_B} \mathcal{L}(F, G, D_A, D_B). \tag{6}$$

**Residual Generator.** The architecture of generator *F* and *G* consists of 4 convolution layers with stride-2, 8 sequentially connected residual blocks, and 4 transposed convolutions to restore the original image size. All the (transposed) convolution layers are followed by LeakyReLU and instance normalization.

**Discriminator.** The discriminator between real and fake images of a specific domain consists of a two layer fully connected neural network. The non-linearity activation function in between is still LeakyReLU. The discriminator takes the task specific features as input, which is detailed below. We notice that during the adversarial training, the weights in the discriminators are updated, while the feature extractors are fixed to ensure that the transfer happens in the desired dimension.

**Task specific feature extractor.** In our specific problem setting, we find it helpful to use predefined feature extractors, which guided the generator to capture camera brand and DR related information. The three feature extractors are:

1. *Mutual information score between color channels.* According to [14, 15], mutual information between channels can quantitatively capture lateral chromatic aberration information.
2. *Normalized color histogram.* Different camera brands can have different color temperatures of the flashlight or different post-processing or enhancing preferences. We characterize these differences using normalized color histogram.
3. *Deep Features from DR Classifier.* The final set of feature descriptor comes from the classification model *M* trained by the source domain fundus image. Specifically, it is the feature vector of size $1 \times n$ after the average pooling layer in ResNet models in our experiment.

See supplementary materials for a detailed description of these feature extractors.

## 3     Experiments and Results

The two purposes of our experiments are: 1) To quantitatively verify the impact of camera-brand-related domain shift on the performance of typical DR classification deep learning models. 2) To prove that our proposed adaptation model can effectively mitigate the domain shift and increase the classification performance of the model on target domain images.

We perform two sets of experiments: 1) We train a ResNet-50 model *M* with images of source domain camera brand, then directly use *M* to test images of both source and target domains. 2) We train the camera-oriented residual-CycleGAN for images of a target domain camera. We use the residual generator *G* to transform the target domain images, and then test *M* for performance. To further prove the effectiveness of camera-oriented residual-CycleGAN, we perform ablation studies on two SOTA adaptation methods, namely DANN[16] and vanilla CycleGAN[17].

### 3.1 Datasets

**EyePACS dataset.** This publicly available dataset is released by Kaggle's Diabetic Retinopathy Detection Challenge [18]. We identify five camera brands in the dataset. See supplementary materials for a detailed process of camera brand identification and the brand labels for each image. The original split of EyePACS dataset was 35,126 images for training and 53,576 for testing, with five DR severity grades. Our new distribution of DR severity grades and camera brands is listed in **Table 1**. We notice that the images of camera A used for training comes both the original training set and the public part of the testing set, while images of camera A (for testing), B, C, D, and E come from the private part of the testing set.

**Private Dataset.** Our private dataset contains 55,326 historical fundus images collected from over 20 hospitals in China. The camera brand information of each image is available. Our collaborative licensed ophthalmologists graded the images for binary classification of positive (has referable DR) vs. negative (has no referable DR). Referable DR is defined as DR with grade II or above according to [19]. The distribution of DR label and domain is summarized in **Table 2**.

**Table 1.** Label and domain distribution of EyePacs dataset

| Camera Brand | Grade 0 | Grade 1 | Grade 2 | Grade 3 | Grade 4 |
|---|---|---|---|---|---|
| A (training) | 14601 | 1301 | 3237 | 532 | 499 |
| A (testing) | 13334 | 1305 | 2927 | 455 | 532 |
| B | 6628 | 540 | 1445 | 181 | 173 |
| C | 3755 | 457 | 570 | 71 | 65 |
| D | 3633 | 411 | 609 | 85 | 72 |
| E | 3831 | 317 | 709 | 173 | 121 |

**Table 2.** Label and domain distribution of the private dataset

| Camera Brand | No. of negative samples | No. of positive samples |
|---|---|---|
| I (for training) | 26,656 | 12,534 |
| I (for testing) | 3,395 | 1,512 |
| II | 1,282 | 2,403 |
| III | 354 | 440 |
| IV | 380 | 683 |
| V | 912 | 558 |

### 3.2 Implementation Details

We first adjust the black margins (crop the horizontal direction and pad the vertical direction if necessary) to make the image a square. Then we resize the adjusted images to 512×512 resolution and normalize to 0 to 1 as the input of the network.

We use the Adam optimizer with a start learning rate of 1e-4 decaying linearly to 1e-5 over 200 epochs. The parameters $\lambda_1$ and $\lambda_2$ in the overall loss function are experimentally set to 0.2 and 5, respectively. In general, $\lambda_1$ and $\lambda_2$ should be set as large as possible, but not to trigger oscillation of loss during adversarial learning. The convergence is determined by visually check a fixed set of transformed images.

We implement the model with paddlepaddle architecture [20, 21]. The training process is relatively computationally intensive. With four NVIDIA Tesla-P40 GPUs, 200 epochs translate to about 80 hours for our resplitted eyePACS dataset and 120 hours for our private dataset, for each of the five camera brand adaptation tasks. We will publicize the camera brand list of the EyePACS dataset and code for training residual-CycleGAN for easier reproduction of the research.

For experiments performed on the EyePACS dataset (5-class grading task) and on the private dataset (binary classification task), we use quadratic weighted kappa [22] and the area under the receiver operating characteristics curve (AUC) as the evaluation criteria, respectively.

### 3.3 Quantitative verification on the impact of domain shift

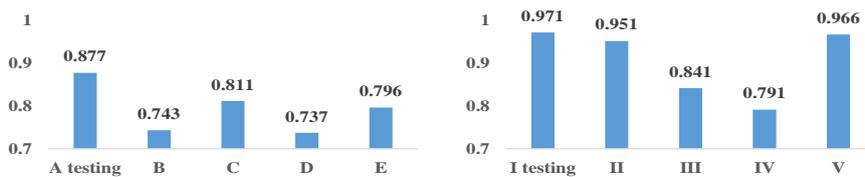

**Fig. 3.** Left: kappa value of the model trained on images taken by camera brand A and tested on various camera brands. Right: AUC value of the model trained on images taken by camera brand I and tested on various camera brands.

**Impact of domain shift in EyePACS dataset.** The performance of the model on testing set of different camera brands is shown in the left chart of **Fig. 3.** The significant performance drop between source domain (camera brand A) and target domains (B, C, D, and E, especially B and D) indicates there is a non-trivial impact from difference in camera brands on the DR classification model.

**Impact of domain shift in private dataset.** Similarly, the right chart of **Fig. 3** shows the drop between source domain (camera brand I) and target domains (especially III and IV), showing that the model is vulnerable to domain shift impact.

We observe several differences between camera brands that correlate with the performance decrease. 1) Overall sharpness: the sharpness of fundus image usually depends on post-processing settings and pixel geometry of the camera. In DR classification, sharpness of edge is an important discriminative feature between hard exudates and drusen. A shift in overall sharpness between source and target domain may confuse the model trained only on one domain. 2) Color distribution: the difference in color distribution or white balance can easily cause the model to confuse dust with a small hemorrhage, leading to the misclassification of the whole image.

### 3.4 Camera-oriented Residual-CycleGAN mitigates the domain shift

**Domain adaptation performance on EyePACS dataset**: The quantitative results are listed in **Table 3**. Our method shows best performance in three out of four target domains and exceeds the model without adaptation by a large margin. This result shows that in this task our method can effectively transform the image to mitigate the domain shift and increase the classification performance on target domains.

**Domain adaptation performance on private dataset**: The results are summarized in **Table 4**. Our method shows best performance in two out of four target domains, which is as effective as DANN.

Besides the quantitative comparison of results, some examples of images in the source domain and target domain, with or without transformation are listed in the supplementary materials for a better qualitative impression.

Compared with the vanilla CycleGAN, our camera-oriented residual-CycleGAN performs better in most of the target domains in the two datasets. We notice that the vanilla CycleGAN performs worse than no adaptation on several of the tasks. Unlike vanilla CycleGAN, our residual-CycleGAN has explicit residual implementation that, combined with the cycle-consistency loss, in effect limits the range of the residue and prevents the generator from creating non-existing lesion or structures. Thus, the residual-CycleGAN almost always outperforms the no adaptation.

The performance of our adaptation approach is better than DANN only by a small margin. However, we notice that DANN, as well as ADDA [23] and other domain invariant feature extraction methods, requires retraining the baseline model when perform the adaptation. In real-world practice the "base model" is an ensemble of many models with complex results merging logic. In these situations, our approach has the advantage of easiness in deployment.

Gulshan et. al. report remarkable AUCs of 0.991 and 0.990 on EyePACS-1 and Messidor-2 testing sets on referable DR detection task [3]. Their training set includes mixed camera brands, presumably covering the testing set brands. As a comparison, our baseline AUC (where training and testing sets are of one identical camera brand) is 0.971. Part of the gap may be due to model architecture difference (Inception-v3 vs. ResNet-50), or suggest that having mixed brands in the training set could assist the model to find robust features detecting referable DR. Since having multiple brands in the training set is quite common in real-world practice, a natural next step is to evaluate and improve the multi-to-one brand adaptation performance.

Table 3. Quadratic weighted kappa for DR grading, EyePACS dataset

| Camera Brand | No adaptation | Ours | DANN [14] | CycleGAN [14] |
|---|---|---|---|---|
| A (testing set) | 0.877 | - | - | - |
| B | 0.743 | 0.740 | 0.764 | **0.782** |
| C | 0.811 | **0.852** | 0.806 | 0.790 |
| D | 0.737 | **0.781** | 0.695 | 0.754 |
| E | 0.796 | **0.804** | 0.780 | 0.729 |

Table 4. AUC for binary DR classification, private dataset

| Camera Brand | No adaptation | Ours | DANN [14] | CycleGAN [14] |
|---|---|---|---|---|
| I (testing set) | 0.971 | - | - | - |
| II | 0.951 | **0.956** | 0.939 | 0.929 |
| III | 0.841 | **0.882** | 0.865 | 0.864 |
| IV | 0.791 | 0.875 | **0.891** | 0.804 |
| V | 0.966 | 0.965 | **0.970** | 0.956 |

## 4 Conclusions

In this paper, our quantitative evaluation show that without domain adaptation, the domain shift caused by different fundus camera brands can significantly impact the performance typical deep learning models, when performing DR classification tasks. Further, our proposed camera-oriented residual-CycleGAN can effectively mitigate camera brands domain shift, improving the classification performance of the original model on target domain images. With our proposed method, the adaptation can be performed directly before a trained classification network, enabling easy deployment.

One future step is to extend the method to a many-to-one or many-to-many setting, i.e. the training or testing set contain different camera brands. A major challenge is defining a proper cycle-consistency loss among multiple domain transformations. The computational cost is also a burden against the scaling of our current method, especially where high resolution is required for detection of certain diseases.